\documentclass[11pt]{ecca}
\usepackage{amsmath}%
\usepackage{amsfonts}%
\usepackage{amssymb}%

\usepackage{graphicx}

\usepackage{tikz}
\usetikzlibrary{shapes}
\usetikzlibrary{fit}

\pgfdeclarelayer{background}
\pgfdeclarelayer{foreground}
\pgfsetlayers{background,main,foreground}

\newtheorem{proposition}{Proposition}

\newtheorem{definition}{Definition}
\newtheorem{example}{Example}

\title{Cooperation in Subset Team Games: Altruism and Selfishness}
\author{Elisha Peterson}
\affiliation{United States Military Academy, West Point, NY 10996-1905\\
Email: \texttt{\footnotesize elisha.peterson@usma.edu}, Phone: \texttt{\footnotesize 845-938-5649}, Fax: \texttt{\footnotesize 845-938-2409}}

\begin{document}
\maketitle

\begin{abstract}
    This paper extends the theory of subset team games, a generalization of cooperative game theory requiring a payoff function that is defined for all subsets of players. This subset utility is used to define both altruistic and selfish contributions of a player to the team. We investigate properties of these games, and analyze the implications of altruism and selfishness for general situations, for prisoner's dilemma, and for a specific game with a Cobb-Douglas utility.
\end{abstract}


\section{Introduction}\label{s:intro}

\begin{quote}
    \textit{
        ``At the interface between [game theory and social choice theory] we find ourselves in a bare landscape dominated by a single large question: what is required to unite the rationality of the individual and the rationality of the group?'' --Michael \cite{bacharach2001}
    }
\end{quote}

Classical cooperative game theory as introduced by \cite{vonneumann1928} is typically used to analyze situations such as oligopolies, when several companies work together to increase their own individual profits. In contrast, many forms of cooperation involve altruism, in which individuals subjugate their interests to those of the group. Consider the user-written encyclopedia \emph{Wikipedia}, which relies on contributions of thousands of volunteers, who often gain little individual benefit from their contributions. In the military, individuals are systematically trained to suppress their individual desires for the good of the group. The golden rule encourages cooperation based upon another's good.

Applying cooperative game theory to such situations requires some mental gymnastics, particularly the argument that every individual has a hidden `utility' that drives every single behavior. While there is truth to this approach, in reality individuals must balance several different utilities, and often make choices to improve a group utility rather than an individual utility, as shown experimentally by \cite{colman2008}. In the natural world, several instances of this kind of group behavior have been collected by \cite{dugatkin1999}.

This paper extends the theory of \emph{subset team games}, and demonstrates their usefulness as a unifier of individual and group utilities. This framework was first defined by \cite{arneypeterson2008}, and builds upon von Neumann's theory while still leaving room for team-oriented aspects of cooperation. Our principle tool is the notion of \emph{cooperation space}, which expands the traditional concept of a \emph{marginal contribution} in classical cooperative game theory into two dimensions. Given a \emph{subset utility function}, as defined in section \ref{s:stgames}, one may compute metrics of altruism $a_A$ and selfish contribution $c_A$ for each subset of players $A$. The location of the coordinate $(a_A,c_A)$ in cooperation space has dramatically different implications for the team. Behaviors in Quadrant I provide the most stability and represent true teamwork, while those in Quadrant II are often found in antagonistic entities such as oligopolies or two-party political systems.

A related approach was pioneered by \cite{bacharach1999}, who introduced the idea of \emph{frames} in the early 1990s to overcome the difficulties in traditional game theory created by \emph{Prisoner's Dilemma} and the \emph{Hi-Lo Game}. His work was posthumously collected in \cite{bacharach2006}, and has since been furthered by \cite{sugden2008} and others. Frames provide a way to reconcile individual and group utilities, allowing for the possibility of \emph{team reasoning} at the individual level. Individuals applying this behavior compute the optimal strategy for the team, and behave as required by that strategy.

We begin with a review of cooperative game theory in section \ref{s:background}, and then go on to define subset team games and explore their properties in section \ref{s:stgames}. The highlights of this section include metrics for altruism and selfish contributions (Definition \ref{d:metrics}), the idea of a \emph{cooperation space}, and additivity properties. The remainder of the paper is devoted to two applications of the theory. Section \ref{s:dilemma} applies the framework to Prisoner's Dilemma, while section \ref{s:cobb} contains an extended quantitative analysis of a Cobb-Douglas utility.

\section{Classical Cooperative Game Theory}\label{s:background}

Our framework is an extension of von Neumann's cooperative game theory (\citeyear{vonneumann1928}). The book by \cite{vonneumann1944} is more comprehensive, while a more modern treatment of this classic material is given by \cite{myerson1997}.

\subsection{TU games}

A \emph{cooperative game with transferable utility} (TU game), sometimes called a \emph{coalition game}, consists of (i) a set of players $T$ and (ii) a \emph{payoff function} (or \emph{utility function}) $u:2^T\to R$ associating a particular value or utility to each subset of $T$.

Note that $2^T$ is the collection of all subsets of $T$. A specific subset $S\subset T$ is called a \emph{coalition}, and $u(S)$ is interpreted as the maximum payoff obtained when these players work together. Roughly speaking, transferable utility indicates that each player has the same value system, so that $u(S)$ may be partitioned among the players.

The \emph{marginal contribution} of a player $a\not\in S$ to a coalition $S$ is
    \begin{equation}\label{eq:marginal}
    m_a(S) \equiv u(S\cup\{a\}) - u(S).
    \end{equation}
This measures the difference in utility of outcome with and without the player $a$. It is easily generalized to marginal contributions $m_A(B)=u(A\cup B)-u(B)$ between disjoint subsets.

\subsection{Allocations}

Assume the team $T$ consists of $n$ players denoted $1,2,\ldots,n$.
An \emph{allocation} $\phi\in R^n$ is a division of the value $u(T)$ earned by the entire population to its players with $\sum_i \phi_i = u(T)$. The \emph{Shapley value} is the important case defined by
    \begin{equation}\label{eq:shapley}
        \phi^{SV}_i = \sum_{S\subset T\setminus\{i\}} \tfrac{|S|!(n-1-|S|)!}{n!} m_i(S\cup\{i\})
                    = \frac1n \sum_{k=0}^{n-1} \tfrac{1}{\tbinom{n-1}{k}}
                              \sum_{\substack{S\subset T\setminus\{i\}\\ |S|=k}} m_i(S\cup\{i\}),
    \end{equation}
where $\binom{n-1}{|S|}=\frac{(n-1)!}{|S|!(n-1-|S|)!}$ is the number of ways to select a subset of size $|S|$ out of $T\setminus\{i\}$. Thus, each player is compensated in proportion to their average marginal contribution. With respect to certain axioms, the Shapley value can be said to be the ``fairest'' possible allocation.

The \emph{core} of a TU game is the set of allocations $\phi\in R^n$ such that $\sum_{i\in S}\phi_i\ge u(S)$ for all coalitions $S\subset T$. Under an allocation in the core, no subset of $T$ could ``defect'' and obtain a better payoff. When the core is nonempty, all players have reason to participate. The Shapley value is not always in the core, and the core may not even exist in some cases. However, the cases in which it is in the core can be partially classified.

A utility function $u$ is \emph{convex} if
    \begin{equation}
        m_i(S\cup\{i\}) \leq m_i(S'\cup\{i\})
    \end{equation}
for all $S\subset S'$ and $i\not\in S$. This means that marginal contributions increase weakly with the size of the coalition. It is not too difficult to show that if a TU game is convex, then the Shapley value is in the core of the game.

A TU game is said to be \emph{superadditive} if $u(A\cup B)\geq u(A)+u(B)$, or equivalently $m_A(B)\geq u(A)$ for all disjoint pairs $A,B$.
In this case, $m_i(S)\ge u(\{i\})$, and so $\phi^{SV}_i\ge u(\{i\})$ for all $i$. This means the Shapley value is \emph{individually rational}.

\subsection{NTU games}

A \emph{cooperative game with non-transferable utility} (NTU game) consists of (i) a set of players $T$, (ii) a set $X$ of possible outcomes, (iii) a \emph{consequence function} $V:2^T\to X$ assigning an outcome to each coalition $S\subset T$, and (iv) a \emph{payoff} or \emph{utility function} $u_a:X\to R$, defined for each player $a\in T$, that associates a value to each possible outcome.

\section{Subset Team Games}\label{s:stgames}

Subset team games generalize cooperative games by allowing each subset of players to have a different assessment of the value of an outcome. The added complexity of the setup allows the marginal contribution to be partitioned into a ``selfish contribution'' and an ``altruistic contribution.'' Definitions \ref{d:stgame} and \ref{d:metrics} were first published by \cite{arneypeterson2008}.

\begin{definition}\label{d:stgame}
    A \emph{subset team game} ($ST$ game) consists of (i) a \emph{team} of players $T$, (ii) a set $X$ of possible outcomes, (iii) a consequence function $V:2^T\to X$ mapping each coalition $S\subset T$ to an outcome, and (iv) a \emph{payoff} or \emph{utility function} $u_A:X\to R$, defined for each subset $A\subset T$.
\end{definition}
The \emph{subset utility} is the function $u_A(S) \equiv u_A(V(S))$. In the context of $u_A(S)$, we call $A$ the \emph{assessing subset}, and $S$ the \emph{coalition}. Frequently it makes sense to assume that $A\subset S$.

\subsection{Altruism and Selfishness}

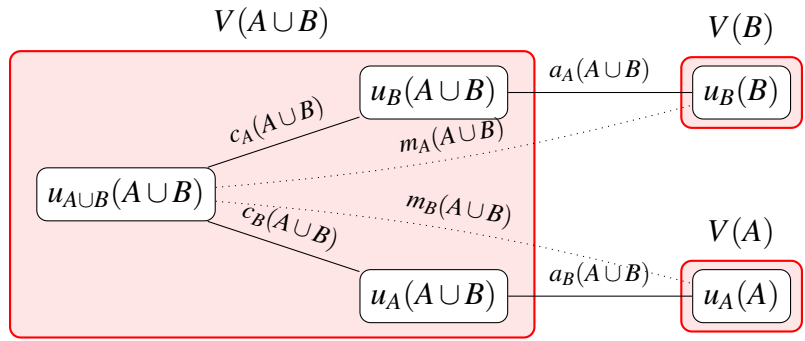
\begin{figure}[htb]
\centering
    \begin{tikzpicture}[grow=right,scale=.9,rounded corners]
        \node[rectangle,draw=black,fill=white](vabab){$u_{A\cup B}(A\cup B)$}[sibling distance=30mm,level distance=45mm]
            child { node[rectangle,draw=black,fill=white](vaab){$u_A(A\cup B)$} child{ node[rectangle,draw=black,fill=white](vaa){$u_A(A)$}
            edge from parent node[sloped,above,scale=.8]{$a_B(A\cup B)$} } edge from parent node[sloped,above,scale=.8]{$c_B(A\cup B)$}  }
            child { node[rectangle,draw=black,fill=white](vbab){$u_B(A\cup B)$} child{ node[rectangle,draw=black,fill=white](vbb){$u_B(B)$}
            edge from parent node[sloped,above,scale=.8]{$a_A(A\cup B)$} } edge from parent node[sloped,above,scale=.8]{$c_A(A\cup B)$}  };
    \begin{pgfonlayer}{background}
        \node[draw=red,fill=red!10,inner sep=0pt,thick,rectangle,fit=(vabab)(vaab)(vbab),label=90:{$V(A\cup B)$}]{};
        \node[draw=red,fill=red!10,inner sep=2pt,thick,rectangle,fit=(vaa),label=90:$V(A)$]{};
        \node[draw=red,fill=red!10,inner sep=2pt,thick,rectangle,fit=(vbb),label=90:$V(B)$]{};
    \end{pgfonlayer}
    \draw(vabab)to[dotted,bend left=5](vaa)node[pos=.5,scale=.8,sloped,above]{$m_B(A\cup B)$};
    \draw(vabab)to[dotted,bend right=5](vbb)node[pos=.5,scale=.8,sloped,above]{$m_A(A\cup B)$};
    \end{tikzpicture}
\caption{Subset utilities for disjoint subsets, grouped by outcome.}\label{fig:metrics}
\end{figure}

\begin{definition}\label{d:metrics}
    Let $A,B\subset T$ be disjoint coalitions. Given the subset utility $u$, the \emph{total marginal contribution} of $A$ to $A\cup B$ is
        \begin{equation}\label{eq:marginal-subset}
            m_A(A\cup B) \equiv u_{A\cup B}(A\cup B)-u_B(B).
        \end{equation}
    The \emph{competitive contribution} of $A$ to $A\cup B$ is
        \begin{equation}\label{eq:competitive-subset}
            c_A(A\cup B) \equiv u_{A\cup B}(A\cup B)-u_{B}(A\cup B).
        \end{equation}
    The \emph{altruistic contribution} of $A$ to $A\cup B$ is
        \begin{equation}\label{eq:altruistic-subset}
            a_A(A\cup B) \equiv u_B(A\cup B)-u_B(B).
        \end{equation}
\end{definition}
Note that the altruistic contribution compares different \emph{outcomes}, while the competitive contribution compares different \emph{assessments}. It is immediate that
    \begin{equation}
        m_A(A\cup B) = c_A(A\cup B) + a_A(A\cup B).
    \end{equation}
We will use $m_A\equiv m_A(T)$, $c_A\equiv c_A(T)$, and $s_A\equiv s_A(T)$ to denote the values when the participating coalition is the entire team $T$. A schematic of these notions is given in Figure \ref{fig:metrics}.

These metrics permit an analysis of the types of contribution made by any subset of players. We will show that cooperation is maximized when the values of $c_A(B)$ and $a_A(B)$ are as high as possible; zero or negative values indicate instabilities in the group.

\begin{example}
    Suppose that $c_A(A\cup B)=0$ for all disjoint $A,B\subset T$. Then $u_{A\cup B}(A\cup B)=u_B(A\cup B)=u_A(A\cup B)$, implying that the subsets $A$ and $B$ value the outcome $V(A\cup B)$ equally. Therefore, \textit{all} subsets value the outcome equally. Consequently, zero selfish cooperation implies that individual value is meaningless; all parts subjugate their own interests to that of the team.

    So there is a function $u:X\to R$ such that $u=u_S$ for all subsets $S\subset T$, and the game reduces to a $TU$ game.
\end{example}

\subsection{Sensible, Cohesive, and Fully-Cooperative ST Games}

A negative value $c_A(A\cup B)$ is a rather curious condition, implying that $B$ by itself values the outcome $V(A\cup B)$ more than the coalition $A\cup B$ values that outcome. It is possible to define subset utilities that have this property, but they are generally outside the scope of our goals for this paper. We generally assume the following condition:
\begin{definition}
    An $ST$ game is \emph{sensible} if $c_A(A\cup B)\ge0$ for all disjoint $A,B\subset S$.
\end{definition}

\begin{example}
    Suppose that $a_A(A\cup B)=0$ for all disjoint $A,B\subset T$. Then $u_B(A\cup B)=u_B(B)$, meaning both outcomes have equal value to $B$. Therefore, \textit{all} outcomes in which $B$ participates have equal value to $B$. Teamwork is meaningless since no player or group of players will gain anything, or lose anything, by working together.
\end{example}
If some altruistic values are negative, then $u_B(A\cup B)<u_B(B)$, implying the coalition $B$ would do better on its own than with the team, and the group will probably be unstable.

This property of altruism motivates the definition of the $ST$-core:
\begin{definition}
    Given an $ST$ game, a coalition $S\subset T$ is \emph{cohesive} if $a_A(A\cup B)\ge0$ for all disjoint $A,B\subset S$. If all coalitions within a game are cohesive, the $ST$ game is \emph{fully-cooperative}. The \emph{core} of a consequence function $V:2^T\to X$ is the set of utility functions that create a fully-cooperative game.
\end{definition}
To summarize, most $ST$ games of interest will be sensible, and games in which teams are likely to be stable are fully-cooperative.

\subsection{Cooperation space}\label{s:coopspace}

Given a subset utility $u$ and a subset $A\subset T$, the pair $(a_A,c_A)$ represents a point in what we call \emph{cooperation space}. Figure \ref{fig:coopspace} shows this space along with the regions marked by \emph{sensibility} and \emph{cohesiveness}. The marginal contribution $m_A$ is positive above the diagonal line $a_A+c_A=0$.

\begin{figure}[htb]
\centering
    \begin{tikzpicture}
        \draw[fill=red!20,draw=none,semitransparent](-2,0)rectangle(2,2);
        \draw[fill=blue!20,draw=none,semitransparent](0,-2)rectangle(2,2);
        \node at(0,1.2){\emph{sensible}};
        \node[rotate=-90] at(1.2,0){\emph{cohesive}};
        \draw[darkgray,dashed](-2.3,-2.3)to[->](2.3,2.3)node[anchor=west]{$m_A$};
        \draw[darkgray,dotted](-2.3,2.3)to(2.3,-2.3)node[sloped,below,pos=.85,scale=.75]{$m_A=0$};
        \draw[thick,->](0,-2.3)to(0,1)(0,1.35)to(0,2.3)node[anchor=west]{$c_A$};
        \draw[thick,->](-2.3,0)to(1,0)(1.35,0)to(2.3,0)node[anchor=south]{$a_A$};
    \end{tikzpicture}
\caption{Cooperation space. Ideal team behaviors take values in Quadrant I.}\label{fig:coopspace}
\end{figure}

An $ST$ game with $n$ players produces $2^n$ points in cooperation space. If these are all in Quadrant I, the game is both sensible and fully-cooperative. This represents ideal teamwork since every subset of players provides benefit both to the team and to other subsets. If some subsets are in Quadrant II, then their behavior decreases the utility of the outcome to some of the players. Quadrants III and IV indicate behaviors arising from non-sensible utilities.

If a subset $A$ has a choice between several different behaviors, one may plot each resulting subset utility in cooperation space. From the team point-of-view, behaviors lying in Quadrant I are the most desirable since they both improve team utility and maintain team stability. Behaviors in Quadrant II with positive marginal contributions are less desirable since team stability is lost. Without external pressure, it is possible the team may break apart because it is rational for certain subsets to do so. Behaviors in Quadrant II have a negative impact on both team utility and team stability. In section \ref{s:cobb}, we use these ideas to assess behaviors in a game defined using a certain Cobb-Douglas utility function.

\subsection{Additivity}

\begin{example}
    Every $NTU$ game with individual utilities $u_a:X\to R$ is also a subset team game. Set $u_A(B)=\sum_{a\in A} u_a(V(B))$. Then
        \begin{equation}\label{eq:additive-competitive}
        c_A(A\cup B)=\sum_{a\in A\cup B}u_a(A\cup B)-\sum_{a\in B}u_a(A\cup B)=\sum_{a\in A}u_a(A\cup B)=u_A(A\cup B).
        \end{equation}
    Therefore, the competitive contributions are precisely the group utilities, and the game is sensible when the utilities are all positive.
\end{example}

In additive games, the value of an outcome is partitioned among the different players. As hinted above, additive $ST$ games can also be used to define $NTU$ games.
\begin{definition}\label{d:additive}
    A subset utility $u_A(S)$ is \emph{additive} or \emph{linear} if
        \begin{equation}
            u_A(S)=\sum_{a\in A}u_{a}(S)
        \end{equation}
    for all $A\subset S\subset T$.
\end{definition}


\begin{proposition}\label{p:additive}
    If a subset utility $u_A(S)$ is additive, then
    (i) $c_A(A\cup B)=u_A(A\cup B)=\sum_{a\in A}u_a(A\cup B)$;
    (ii) $a_A(A\cup B)=\sum_{b\in B}\left(u_b(A\cup B)-u_b(B)\right)$;
    (iii) the game is sensible if and only if $u_a(A\cup B)\ge0$ for all $a\in A$; and
    (iv) the game is fully-cooperative if and only if $u_b(A\cup B)\geq u_b(B)$ for all players $b\in B$.
\end{proposition}
Given the relationship between stability and cohesiveness, the fourth statement above implies that a team is stable when each player does better with the addition of the players in $B$.

A similar notion is the following.
\begin{definition}\label{d:coadditive}
    A subset utility $u_A(S)$ is \emph{co-additive} if
        \begin{equation}
            u_A(S)=\sum_{b\in S}u_A(b)
        \end{equation}
    for all $A\subset S\subset T$. It is \emph{bi-additive} if it is both additive and co-additive.
\end{definition}
Co-additivity implies that each subset $A\subset T$ has a unique assessment of the value of the player $b\in B$ to the team, when that player participates. Otherwise put, the outcome is in some sense additive relative to the participants. A coalition $A$ will want any player $b$ with $u_A(b)\ge0$ to join.

\begin{proposition}\label{p:coadditive}
    If a subset payoff function is co-additive, then
    (i) $a_A(A\cup B)=u_B(A)=\sum_{a\in A}u_B(a)$;
    (ii) $c_A(A\cup B)=\sum_{b\in A\cup B}\left(u_{A\cup B}(b)-u_B(b)\right)$;
    (iii) the game is fully-cooperative if and only if $u_B(a)\ge0$ for all $a\not\in B$; and
    (iv) the game is sensible if and only if $u_{A\cup B}(b)\ge u_B(b)$ for all $b\in A\cup B$.
\end{proposition}

\begin{proposition}\label{p:biadditive}
    If a subset payoff function is bi-additive, then
        \begin{equation}
            u_A(S) = \sum_{a\in A,b\in S} u_a(b).
        \end{equation}
    Moreover, $c_A(A\cup B)=u_A(A\cup B)$ and $a_A(A\cup B)=u_B(A)$.

    If a bi-additive subset team game is fully-cooperative and $u_a(a)\ge0$ for all $a\in T$, then it is sensible.
\end{proposition}

In bi-additive games, the subset utility is determined by a $n\times n$ matrix of values, where $n$ is the number of players. The entries of the matrix are individual perceptions of other players' contribution to the team. Alternately, bi-additive $ST$ games may be represented by a weighted graph, where the vertices are players and the edge from $a$ to $a$ is labeled by $u_a(b)$, the perceived value of $b$ to $a$. Then, the competitive contribution of a subset $A$ is $u_A(A\cup B)$, the total weighted degree of edges terminating in $A$. The altruistic contribution of a subset $A$ is $u_B(A)$, the total weighted degree of edges starting in $A$ and terminating outside $A$. See Figure \ref{fig:biadditive}.

\begin{figure}[htb]
\centering
    \begin{tikzpicture}[every node/.style={fill=black,draw=none,circle,inner sep=1pt},scale=2]
        \draw[draw=black,fill=black!10,circle](0,0)circle(.6);\node[rectangle,fill=none]at(0,.45){$A$};
        \draw[draw=black,fill=black!10,circle](2,0)circle(.6);\node[rectangle,fill=none]at(2,.45){$B$};
        \node(aa)at(0:.5){};\node(ab)at(120:.5){};\node(ac)at(240:.5){};
        \begin{scope}[shift={(2,0)}]
            \node(ba)at(20:.5){};\node(bb)at(120:.5){};\node(bc)at(220:.5){};\node(bd)at(320:.5){};
        \end{scope}
        \begin{scope}[gray,thick,dotted,->]
            \draw(ba)to(bc);\draw(bb)to(bc);\draw(bc)to[bend left=20](bd);\draw(bd)to[bend left=20](bc);
        \end{scope}
        \begin{scope}[blue,thick,->]
            \draw(aa)to(ab);\draw(ac)to(ab);\draw(ac)to[bend left=20](aa);\draw(aa)to[bend left=20](ac);
            \draw(ba)to[bend right](ab);\draw(ba)to[bend right=10](aa);\draw(bb)to[bend left=20](ac);
        \end{scope}
        \begin{scope}[red,thick,dashed,->]
            \draw(aa)to[bend right=10](ba);\draw(aa)to[bend right](bc);\draw(ac)to[bend right](bc);\draw(ac)to[bend right](bd);
        \end{scope}
        \node[blue,rectangle,fill=none]at(-.3,.8){$c_A(A\cup B)$};
        \node[red,rectangle,fill=none]at(1,-1){$a_A(A\cup B)$};
    \end{tikzpicture}
\caption{Altruistic and competitive cooperation for a bi-additive subset team game.}\label{fig:biadditive}
\end{figure}

\section{Prisoner's Dilemma}\label{s:dilemma}

This section applies subset team games to a simple scenario. \emph{Prisoner's Dilemma} is a two-player game in which players are being interrogated, and must decide whether to keep silent (cooperating) or turn in their partner (defecting). A representative payoff matrix follows.
    $$
    \tikz{
        \draw(0,0)rectangle node{3,0}(1,1);
        \draw(1,0)rectangle node{1,1}(2,1);
        \draw(0,1)rectangle node{2,2}(1,2);
        \draw(1,1)rectangle node{0,3}(2,2);
        \node[anchor=east]at(-.1,.5){def};
        \node[anchor=east]at(-.1,1.5){coop};
        \node[anchor=south]at(.5,2.1){coop};
        \node[anchor=south]at(1.5,2.1){def};
        }
    $$

The dilemma arises because the \emph{rational choice} for both players is to defect. On the other hand, they would do better if both chose to cooperate. In social experiments, real players often do choose to cooperate, as shown in a number of studies collected by \cite{sally1995}. There are a number of ways to resolve this paradox.
\begin{itemize}
    \item \textbf{Lack of Information.} In real life, players do not always know what is in their best interest.
    \item \textbf{Misrepresentation of Values.} The utility given does not represent the player's actual best interests.
    \item \textbf{Iteration and Trust.} If the game is repeated many times, it is rational for players to develop trust and cooperate with those they trust (an approach studied extensively by \cite{axelrod1984}).
\end{itemize}

From the viewpoint of subset team games, it makes sense to consider players as either working together as a ``team'' or working separately. Each situation provides a different value to the players, as follows:
\begin{center}
\begin{tabular}{rccc}
    utility & $V(A,B)$ & $V(A)$ & $V(B)$ \\
    $u_{A,B}$ & 4 & - & - \\
    $u_A$ & 2 & 1 & - \\
    $u_B$ & 2 & - & 1 \\
\end{tabular}
\end{center}
Individually, the best a player can hope to receive is 1. But if they work as a team, their individual utility goes up to 2. This means that each player's altruistic contribution is 1. The team's overall utility is 4, so each player's selfish contribution is 2.

Therefore, the team utility occupies Quadrant I of cooperation space; the game is both sensible and fully-cooperative. Since the altruistic contributions are positive, teamwork is ``robust'' in the sense that working apart gives a worse outcome. The conditions are right for teamwork to flourish. This does not mean that an individual cannot stand to gain by defecting, merely that defecting provides a poor long-term strategy.

\section{Teamwork with a Cobb-Douglas Utility Function}\label{s:cobb}

We now consider a more complex scenario. There are several ways to assess the value of outcomes consisting of two measurable quantities that are both desirable. For $0\le\theta\le1$, the Cobb-Douglas utility function
    \begin{equation}
    U_\theta(y,z)=y^\theta z^{1-\theta}
    \end{equation}
puts the highest value on outcomes that balance two positive quantities $y$ and $z$. In particular, when $y+z$ is fixed, elementary calculus can be used to show that the maximum value for $U_\theta(y,z)$ occurs when $y=\left(\frac{\theta}{1-\theta}\right)z$.


We consider an $ST$ game in which each player in $T$ has a fixed pool of resources to contribute. The combined contributions of the participating players determines the value of the outcome, which is in turn divided up among the players. This section follows an approach similar to that of \cite{axtell2002}, who evaluated many of the same concepts from a different point-of-view. The main difference is the focus here on altruism, selfish contributions, and cooperation space.

\subsection{Definitions}
Given a player $a\in T$, let $X_a$ denote that player's resources, and $x_a\le X_a$ their contributions. Let $\hat x_a=X_a-x_a$ represent the amount of resources kept in reserve. Given a subset $S\subset T$, define $\hat x_S=X_S-x_S$ similarly.

Let $f(x_S)$ be the total value of the outcome given the contribution $x_S$ by a subset $S\subset T$, let $f_a(S)\ge0$ denote the resulting payment to a player $a\in S$, and let $f_A(S)=\sum_{a\in A}f_a(S)$ denote the payment to a subset $A\subset S$. We require $f_S(S)\le f(x_S)$, since $f(x_S)$ is the maximum value that can be distributed among the players.

One obtains an $ST$ game with subset utility
    \begin{equation}\label{eq:cd-utility}
    u_A(S)=U_\theta(f_A(S),\hat x_A)=\left(f_A(S)\right)^\theta\left(\hat x_A\right)^{1-\theta}.
    \end{equation}
This means that the subset $A$ wishes to balance its payment $f_A(S)$ with its reserves $\hat x_A$.

The game is in general neither additive nor co-additive. However, it is always sensible:

\begin{proposition}\label{p:cd-utility}
    The subset utility $U_A(S)$ defined in \eqref{eq:cd-utility} is always sensible. It is fully-cooperative if and only if
    $f_B(A\cup B)\ge f_B(B)$.
\end{proposition}
Intuitively, the fully-cooperative condition means that as the pool of resources increases, so does each player's payoff.

\subsection{Payoff Schemes}
For the remainder of the section, assume that $f$ is continuous and differentiable with $f(0)=0$. We will investigate three different total allocations $f(x_S)=\sum_{a\in S}f_a(x_S)=f_S(S)$, defined as follows:
\begin{itemize}
    \item $f_A(S) = \frac{x_A}{x_S}f(x_S)$ (\emph{proportional});
    \item $g_A(S) = \frac{|A|}{|S|}f(x_S)$, where $|A|$ is the number of players in the subset $A$ (\emph{equal});
    \item $h_A(S) = \gamma f_A(S) + (1-\gamma) g_A(S)$ for $0\le\gamma\le1$ (\emph{hybrid}).
\end{itemize}
Since $f_S(S)=g_S(S)=h_S(S)=f(x_S)$, all three payoffs divide the entire value achieved by $S$. As \cite{axtell2002} mentions, proportional payoffs are ideal but somewhat impractical since it requires perfect knowledge of player's contributions and resources. On the other hand, equal payoffs are ripe for exploitation.

\begin{figure}[htb]
\centering
\includegraphics*[scale=.45]{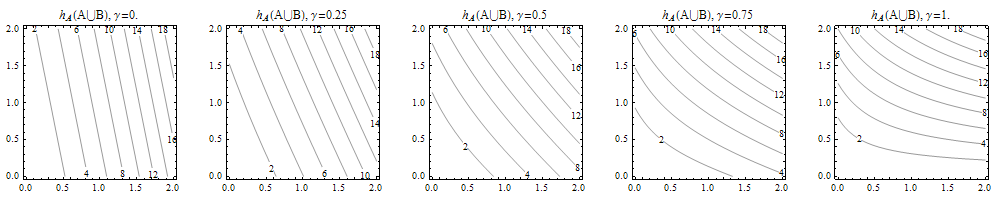}
(a)
\includegraphics*[scale=.45]{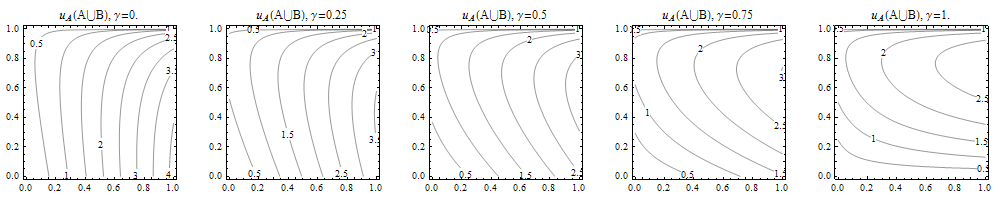}
(b)
\caption{Hybrid payoffs (a) and utilities (b) for a subset $A$ of players $A\cup B$ given the value function $f(x)=x^{1.5}$, as a function of the average contributions of players in disjoint subsets of size $|A|=2$ (horizontal axis) and $|B|=10$ (vertical axis).}\label{fig:payoffs}
\end{figure}

Figure \ref{fig:payoffs}a shows the payoffs $h_A(A\cup B)$ for $f(x)=x^{1.5}$ and various values of $\gamma$. Figure \ref{fig:payoffs}b shows the subset utility $u_A(A\cup B)$ for $\theta=0.75$, indicating a significant preference for reward over reserve resources. The leftmost graphs indicate equal payoffs, and the rightmost indicate proportional payoffs.


\subsection{Team Stability}
\cite{axtell2002} notes that equal and hybrid payoffs may give rise to nonstable teams. One obtains the same conclusion by considering cooperation space. The game is stable when rational individual and team strategies lie in Quadrant I of cooperation space, as mentioned in section \ref{s:coopspace}. By Proposition \ref{p:cd-utility}, this occurs when $f_B(A\cup B)\geq f_B(B)$.

In the proportional payoff case, this means $\frac{x_B}{x_A+x_B}f(x_A+x_B)\ge f(x_B)$. This condition is always satisfied by a function with
    \begin{equation}
    \frac{f(y)}{y}\ge\frac{f(x)}{x}
    \end{equation}
for $y\ge x$, which implies $\left(\frac{f(x)}{x}\right)'\ge0$ or $f'(x)\ge\frac{f(x)}{x}$. So the rate of return on contributions increases with the size of the contributions.

For equal payoffs, one may rewrite the condition $\frac{|B|}{|A|+|B|}f(x_A+x_B)\ge f(x_B)$ as
    \begin{equation}
    \frac{f(x_S)}{|S|}\ge\frac{f(x_B)}{|B|},
    \end{equation}
where $S=A\cup B$. So the average return must increase with the size of the group. The only nonnegative function that can satisfy this property universally is $f=0$, since the left side may be made arbitrarily small by increasing $|A|$ while keeping $x_A$ fixed. Large groups can only be cohesive if their contributions are analogously large. In the case $|B|=1$, the inequality simplifies to $f(x_b)\le\frac{f(x_S)}{|S|}$, so a single player's contribution is limited by the average value achieved. A player that contributes more than his share will be tempted to leave the group.

The same analysis holds in the hybrid case, although the formulas become more complicated. Setting $A\cup B=S$ and $|B|=1$ again, the condition in Proposition \ref{p:cd-utility} becomes
    \begin{equation}
    \left(\gamma\frac{x_b}{x_S}+(1-\gamma)\frac{1}{|S|}\right)f(x_S)\ge f(x_b).
    \end{equation}
Again, no function $f\neq 0$ will work universally (unless $\gamma=1$), and the group size is limited, although to a lesser extent, by the size of individual contributions.

\subsection{A Specific Case}

\begin{figure}
\centering
\includegraphics*[scale=.5]{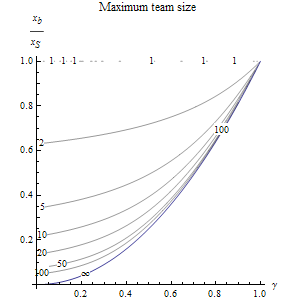}
\caption{The maximum stable team size as a function of $\gamma$ and $\frac{x_b}{x_S}$. Below the curves, all team sizes are stable.}\label{fig:size}
\end{figure}

We will now focus on the specific case of a power function $f(x)=\alpha x^\beta$ with $\alpha>0$ and $\beta>1$. Such a function satisfies $f'(x)\ge\frac{f(x)}{x}$, and so leads to a fully-cooperative game in the case of proportional payoffs. In the equal payoff case, the group size condition reduces to $|S|\le\left(\frac{x_S}{x_b}\right)^\beta$. In the hybrid payoff case, the group size condition reduces to
    \begin{equation}
    |S|\le\frac{1-\gamma}{\left(\frac{x_b}{x_S}\right)^\beta-\gamma\frac{x_b}{x_S}}
    \end{equation}
if the denominator is positive. (If the denominator is negative, then $|S|$ may be taken to be arbitrarily large.) The maximum stable team sizes for various values of $\gamma$ and $\frac{x_b}{x_S}$ is shown in Figure \ref{fig:size}.

\begin{figure}
\centering
\includegraphics*[scale=.41]{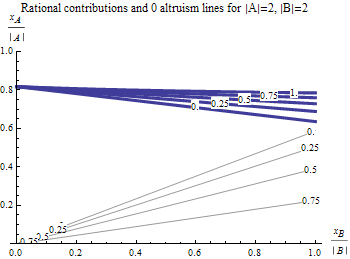} \includegraphics*[scale=.41]{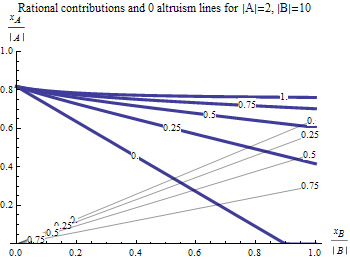} \includegraphics*[scale=.41]{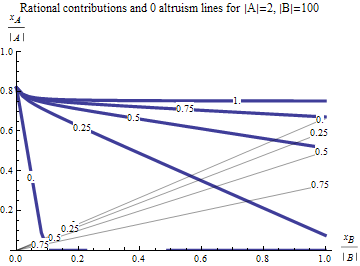}
\caption{Rational choices for $A$ (thick lines) as a function of average contributions for $B$, overlaid with zero altruism lines, for various values of $\gamma$ and $|B|$.}\label{fig:altutility}
\end{figure}

From an economic standpoint, players are apt to make \emph{rational} choices, choosing to contribute an amount that maximizes their own utility. Will this produce a stable result? Figure \ref{fig:altutility} shows rational contributions, as a function of the average contribution of a disjoint subset $B$, overlayed with lines of 0 altruism $a_A(A\cup B)=0$. If a contribution is above the line, the coalition is cohesive. When the lines for some $\gamma$ intersect, however, there is a range of values where an individual's optimal contribution leads to a non-cohesive team. In such cases, players have the opportunity to ``cheat'' or ``free-ride'' by taking advantage of near-equal payoffs.

\begin{figure}
\centering
\includegraphics*[scale=.37]{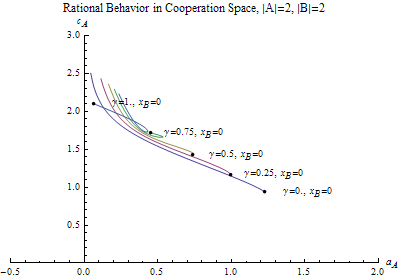} \includegraphics*[scale=.37]{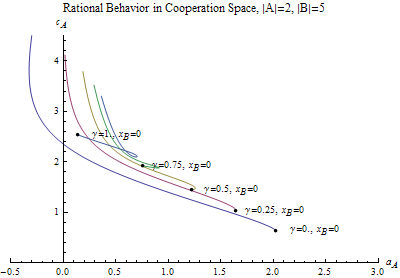} \includegraphics*[scale=.37]{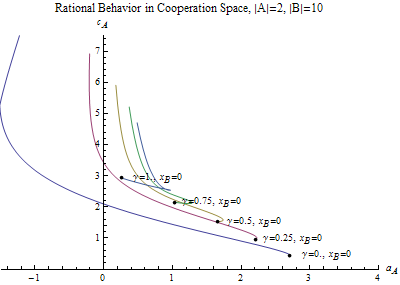}
\caption{Paths of rational behavior in cooperation space. The curves have parameter $\frac{x_B}{|B|}\in[0,1]$, and each curve represents a different value of $\gamma$.}\label{fig:coopspace-cobb}
\end{figure}

We can also examine the game in \emph{cooperation space}. Figure \ref{fig:coopspace-cobb} shows curves with a single parameter representing the average contribution $\frac{x_B}{|B|}$. The behavior of subset $A$ is the rational behavior as indicated in Figure \ref{fig:altutility}. Multiple paths are plotted, for values of $\gamma$ between 0 and 1, and each plot shows a different value of $|B|$.
These plots capture in a single figure the value of the subset team game framework. One can make the following observations:
\begin{itemize}
\item When the team sizes are equal (first graph), the game is always fully-cooperative; the worst case for stability occurs when $|B|$ is much larger than $|A|$ (third graph).
\item Team stability is greater for equal payoffs ($\gamma=0$) when $\frac{x_B}{|B|}$ is small, but greater for proportional payoffs ($\gamma=1$) in general. This is seen by comparing the starting locations of the curves ($\frac{x_B}{|B|}=0$) with their endpoints ($\frac{x_B}{|B|}=1$).
\item The marginal contributions associated with rational behavior increase as $\gamma$ approaches 1.
\item As $\frac{x_B}{|B|}$ approaches $1$, the value of ``cheating'' increases significantly for small $\gamma$. This indicates that the players in $A$ are taking advantage of the larger contributions of $B$.
\end{itemize}
In short, the payoff scheme can have dramatic implications for the cohesivity of the team, ensuring that players are not marginalized.

\section{Conclusion}\label{s:conclusion}
The subset team game framework unifies individual and team utilities, and is general enough to apply to a wide variety of scenarios. Perhaps the most important contribution is the idea of cooperation space and the specific metrics of altruism and selfish cooperation. This extends the marginal contribution $m_A$ to a pair $(a_A,c_A)$, as visualized in Figures \ref{fig:coopspace} and \ref{fig:coopspace-cobb}. The applications of this viewpoint are endless. One could promote team stability and optimize a team's success by changing the payoff scheme to increase the level of altruism; one could identify ``cheaters'' within a system by investigating the behavior of individual or subsets in this space; one could identify situations in which additional incentives are required to encourage individuals to work together.

These ideas might also be applied to evolutionary models of team formation, with teams allowed to form when individual behaviors are grouped together in Quadrant I of cooperation space. It is an open and interesting question whether teams that adopt rules forcing constituents to have some degree of altruism are more successful in the long run than teams with no such rules. In one sense, it is required to ensure team stability. For example, two-party political systems tend to live in Quadrant II since they have opposing viewpoints, and consequently often find it very difficult to pass certain kinds of legislation.

The difficulty in practice comes in finding suitable functions for subset utilities, which require a significant amount of information. This paper has shown how the process might work in the case of balancing resource reserves, using a Cobb-Douglas function. The ideal real-world solution is to balance equal and proportional payoffs, perhaps using estimates of player contributions rather than precise values. The analysis in this paper provides one step towards understanding dynamic groups in the presence of individual goals.


\appendix
\section{Proof of Propositions \ref{p:additive} and \ref{p:coadditive}}
    Condition (i) follows from \eqref{eq:additive-competitive} and (ii) follows from the definition of altruistic contribution.
    Then (i) implies (iii) and (ii) implies (iv). The proof of Proposition \ref{p:coadditive} is similarly straightforward.

\section{Proof of Proposition \ref{p:biadditive}}
    The first statement follows immediately from Propositions \ref{p:additive} and \ref{p:coadditive}.

    For the second statement, note that $c_A(A\cup B)=u_A(A\cup B)=u_A(A)+u_A(B)$. The fully-cooperative condition implies $u_A(B)\ge0$, while together with $u_a(a)\ge0$ it implies $u_A(A)\ge 0$. Therefore, the two conditions imply $c_A(A\cup B)\ge 0$, which is the condition for sensibility.

\section{Proof of Proposition \ref{p:cd-utility}}

Since $\hat x_{A\cup B}\ge\hat x_B$ and $f_{A\cup B}(A\cup B)\ge f_B(A\cup B)$, it follows that
    \begin{equation}
        c_A(A\cup B)=\left(f_{A\cup B}(A\cup B)\right)^\theta\left(\hat x_{A\cup B}\right)^{1-\theta}
            -\left(f_B(A\cup B)\right)^\theta\left(\hat x_B\right)^{1-\theta}
        \ge 0.
    \end{equation}
Therefore, the game is always sensible.

The altruistic contribution is
    \begin{multline}
        a_A(A\cup B)=\left(f_B(A\cup B)\right)^\theta\left(\hat x_B\right)^{1-\theta}
            -\left(f_B(B)\right)^\theta\left(\hat x_B\right)^{1-\theta}\\
            =\left(\left(f_B(A\cup B)\right)^\theta-\left(f_B(B)\right)^\theta\right)\left(\hat x_B\right)^{1-\theta}.
    \end{multline}
So the game is fully-cooperative if and only if $f_B(A\cup B)\ge f_B(B)$ for all disjoint $A,B$.

\section*{Acknowledgments}
This work is supported by an NRC/Davies Postdoctoral Research Fellowship under the guidance of Chris Arney at the Army Research Office. I have benefitted greatly by conversations with Chris Arney, Lucas Gebhart, Joshua Lospinoso, and Andrew Plucker regarding this work.

\bibliographystyle{ecca}
\bibliography{games-bib}

\end{document}